# Dataset Growth in Medical Image Analysis Research

**Yuval Landau and Nahum Kiryati**

School of Electrical Engineering, Tel Aviv University, Tel Aviv, Israel

Corresponding author: Nahum Kiryati (nk@eng.tau.ac.il).

We thank Shoham Rochel for reviewing some of the raw data and Aya Vituri of the Statistical Consulting Laboratory at Tel Aviv University for her advice. This study was supported in part by the Blavatnik Interdisciplinary Cyber Research Center at Tel Aviv University. Nahum Kiryati is the incumbent of the Manuel and Raquel Klachky Chair of Image Processing at Tel Aviv University.

**ABSTRACT** Medical image analysis studies usually require medical image datasets for training, testing and validation of algorithms. The need is underscored by the deep learning revolution and the dominance of machine learning in recent medical image analysis research. Nevertheless, due to ethical and legal constraints, commercial conflicts and the dependence on busy medical professionals, medical image analysis researchers have been described as "data starved". Due to the lack of objective criteria for sufficiency of dataset size, the research community implicitly sets ad-hoc standards by means of the peer review process. We hypothesize that peer review requires researchers to report the use of ever-increasing datasets as one condition for acceptance of their work to reputable publication venues. To test this hypothesis, we scanned the proceedings of the eminent MICCAI (Medical Image Computing and Computer-Assisted Intervention) conferences from 2011 to 2018. From a total of 2136 articles, we focused on 907 papers involving human datasets of MRI (Magnetic Resonance Imaging), CT (Computed Tomography) and fMRI (functional MRI) images. For each modality, for each of the years 2011-2018 we calculated the average, geometric mean and median number of human subjects used in that year's MICCAI articles. The results corroborate the dataset growth hypothesis. Specifically, the annual median dataset size in MICCAI articles has grown roughly 3-10 times from 2011 to 2018, depending on the imaging modality. Statistical analysis further supports the dataset growth hypothesis and reveals exponential growth of the geometric mean dataset size, with annual growth of about 21% for MRI, 24% for CT and 31% for fMRI. In slight analogy to Moore's law, the results can provide guidance about trends in the expectations of the medical image analysis community regarding dataset size.

## I. INTRODUCTION

Medical image analysis is an active research field, focusing on computational methods for the extraction of clinically-useful information from medical images. Research in medical image analysis critically depends on the availability of relevant medical image sets (datasets) for tasks such as training, testing and validation of algorithms. It is widely accepted that solid medical image analysis research requires the use of sufficiently large datasets. This notion is rooted in classical statistical estimation [1] and classification [2] theories.

Obtaining relevant medical images is challenging and costly, as it requires cooperation of medical professionals and institutes, and alleviation of ethical, legal and often commercial conflicts. For example, making a clinical medical image available for research usually requires a collaborating clinical expert to obtain regulatory approvals, to find a relevant image in an institutional archive, to interpret the image and to remove identifying details. Thus, for research in medical image analysis, relevant images are usually in high demand and short supply. In a recent conference [3], "the common theme from attendees was that everyone participating in medical image evaluation with machine learning is data starved." Thus, typical dataset sizes in medical image analysis research are a far cry from the common perception of big data in healthcare [4].

Since 2012, the Deep Learning paradigm revolutionized the field of medical image analysis [5], underscoring the significance of large image datasets. Thus, researchers are torn between the need for large image datasets, i.e., datasets containing medical images of many subjects, and the cost and effort required to obtain them.

In contrast to classical estimation, detection and classification problems, where sample size planning may yield to analysis [1-2, 6-10], theoretical understanding of deep learning is limited. The trend is that increasing the training dataset size improves the performance of deep learning networks [11]. Small datasets are associated with overfitting and poor generalization performance on unseen data [11]. In the presence of rare pathologies, small datasets are known to result in class imbalance and inadequate training [3]. Yet, solid objective criteria for dataset size are difficult to obtain.

TABLE 1
MICCAI SUBMISSION AND ACCEPTANCE DATA (MAIN CONFERENCE)

|      | Submitted papers | Accepted papers | Acceptance rate |
|------|------------------|-----------------|-----------------|
| 2011 | 819              | 251             | 30%             |
| 2012 | 781              | 252             | 32%             |
| 2013 | 798              | 262             | 33%             |
| 2014 | 862              | 253             | 29%             |
| 2015 | 810              | 263             | 32%             |
| 2016 | 756              | 228             | 30%             |
| 2017 | 800              | 255             | 32%             |
| 2018 | 1068             | 372             | 35%             |

In practice, since dataset size is associated with research quality, peer review processes of reputable publication venues implicitly set ad-hoc thresholds on dataset size. Thus, a manuscript is accepted for publication or presentation in a peer-reviewed venue only if the dataset size is regarded by the referees as sufficiently large. Consequently, dataset sizes appearing in published articles reflect the implicit standard for dataset size. Since the standard is not static, researchers are often dismayed to discover that the dataset used in their current study, where the number of subjects was previously considered sufficiently large, is no longer up to expectations. This uncertainty leads to marginalization and loss of potentially valuable research.

The purpose of this research is to better understand the temporal trends in the implicit community thresholds on acceptable dataset size. We hypothesize that these thresholds grow over time, such that ever increasing datasets are required for acceptance of a manuscript to a reputable venue. Hopefully, our results will provide researchers, funding agencies, program committees and editors with rough guidelines regarding the community standards for dataset size in current and future medical image analysis studies.

## II. METHODS

We reviewed the proceedings of the annual MICCAI conference from 2011 to 2018 [12-19], and noted the numbers of human subjects included in the datasets used. MICCAI, the acronym for Medical Image Computing and Computer Assisted Intervention, is a leading conference in the field, with a rigorous peer-review process (typically at least three reviewers, double-blind). Table 1 shows the number of submitted papers, the number of accepted papers (oral and posters) and the acceptance ratio per year (main conference only, excluding satellite events). We preferred monitoring a conference series rather than an archival journal, since the conference review period is much shorter than that of quality journals, implying a shorter sampling aperture, hence better temporal sampling.

We focused on studies involving three important imaging modalities: Magnetic Resonance Imaging (MRI), Computed Tomography (CT) and Functional MRI (fMRI). Taken together, this selection covers a substantial portion of the research articles in the MICCAI proceedings. Being well-established modalities, a good number of articles associated with each modality is found in each annual edition of the MICCAI proceedings, allowing meaningful statistical study.

We only considered datasets referring to human subjects, rather than animal or other datasets. Given the stringent regulatory framework regarding human data [20], we believe that the challenges and trade-offs associated with collection and use of human medical data are unique and justify exclusion of non-human datasets from this study. Nevertheless, *in-utero* and *post-mortem* human datasets are included. We define the dataset size to be the number of distinct human subjects, rather than the number of test images or similar data structures, as the number of human subjects better reflects the recruitment effort.

For each year and for each modality, we found the relevant articles in the MICCAI proceedings and extracted from each article the dataset size. No distinction was made between data used for training, validation, testing or any other use. With few exceptions, articles associated with several imaging modalities were considered with regard to one of the modalities, with preference to fMRI over CT and MRI. Table 2 shows the number of relevant MICCAI articles, i.e. the articles included in our analysis, per year and imaging modality. The growth in 2018 corresponds to the larger overall number of articles in MICCAI 2018 (see Table 1).

To obtain descriptive results (Section III), for each year and modality the average, geometric mean and median dataset sizes were calculated. In computing the average, to reduce the effect of outliers, the largest value and smallest value were discarded.

Statistical analysis of the data (Section IV) was performed, for each modality separately, using SPSS v.25 and R v.3.6.0. p-values were corrected for multiple comparisons using the Benjamini-Hochberg (BH) procedure, with $p<0.05$ considered significant.

TABLE 2
NUMBER OF ARTICLES INCLUDED IN THE ANALYSIS, PER YEAR AND IMAGING MODALITY

|      | 2011 | 2012 | 2013 | 2014 | 2015 | 2016 | 2017 | 2018 |
|------|------|------|------|------|------|------|------|------|
| MRI  | 62   | 63   | 76   | 63   | 66   | 69   | 75   | 94   |
| CT   | 36   | 24   | 20   | 40   | 23   | 14   | 30   | 36   |
| fMRI | 11   | 10   | 14   | 10   | 14   | 16   | 15   | 26   |

TABLE 3
AVERAGE, GEOMETRIC MEAN AND MEDIAN **MRI** DATASET SIZES
(NUMBER OF SUBJECTS) USED IN MICCAI ARTICLES IN EACH OF
THE YEARS 2011-2018

| **MRI** | 2011 | 2012 | 2013 | 2014 | 2015 | 2016 | 2017 | 2018 |
|---|---|---|---|---|---|---|---|---|
| average | 74.3 | 52.2 | 79.9 | 139.2 | 65.6 | 163.6 | 178.1 | 250.6 |
| geom. mean | 21.7 | 19.2 | 28.5 | 43.0 | 27.4 | 64.2 | 62.5 | 68.6 |
| median | 23 | 20 | 21 | 54 | 33 | 64 | 80 | 67 |

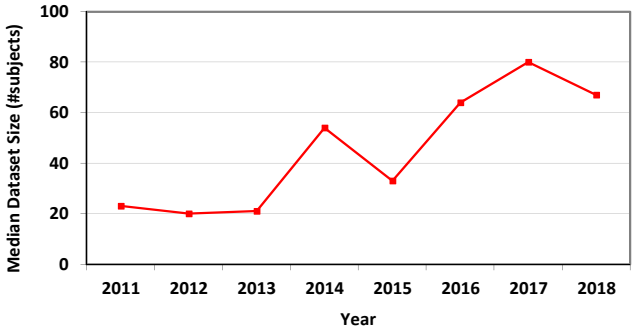

**FIGURE 1.** Median size of datasets used in MICCAI articles related to MRI in each of the years 2011-2018.

## III. DESCRIPTIVE RESULTS

For MRI, the annual average, geometric mean and median dataset sizes, used in MICCAI articles in each of the years 2011-2018, are shown in Table 3. It can be seen that all three measures roughly trebled over the study period. The higher values of the average values follow from the occasional use of large publicly accessible datasets, see e.g. [21], affecting the averages but not the median values. A graphical representation of the median values per year is shown in Figure 1.

TABLE 4
AVERAGE, GEOMETRIC MEAN AND MEDIAN **CT** DATASET SIZES
(NUMBER OF SUBJECTS) USED IN MICCAI ARTICLES IN EACH OF
THE YEARS 2011-2018

| **CT** | 2011 | 2012 | 2013 | 2014 | 2015 | 2016 | 2017 | 2018 |
|---|---|---|---|---|---|---|---|---|
| average | 54.4 | 26.8 | 40.4 | 48.0 | 71.6 | 71.3 | 143.9 | 504.0 |
| geom. mean | 19.1 | 15.6 | 26.4 | 20.3 | 29.3 | 39.7 | 35.7 | 102.0 |
| median | 17 | 16 | 33 | 20 | 29 | 24 | 28 | 72 |

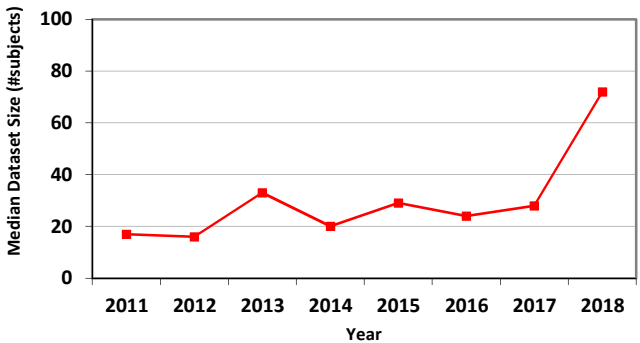

**FIGURE 2.** Median size of datasets used in MICCAI articles related to CT in each of the years 2011-2018.

TABLE 5
AVERAGE, GEOMETRIC MEAN AND MEDIAN *FUNCTIONAL* **MRI**
DATASET SIZES (NUMBER OF SUBJECTS) USED IN MICCAI
ARTICLES IN EACH OF THE YEARS 2011-2018

| **fMRI** | 2011 | 2012 | 2013 | 2014 | 2015 | 2016 | 2017 | 2018 |
|---|---|---|---|---|---|---|---|---|
| average | 21.3 | 31.9 | 32.3 | 67.5 | 86.6 | 111.7 | 151.6 | 264.4 |
| geom. mean | 17.3 | 27.9 | 25.2 | 65.1 | 59.8 | 93.7 | 68.0 | 131.6 |
| median | 15 | 29 | 25 | 64 | 53 | 86 | 46 | 191 |

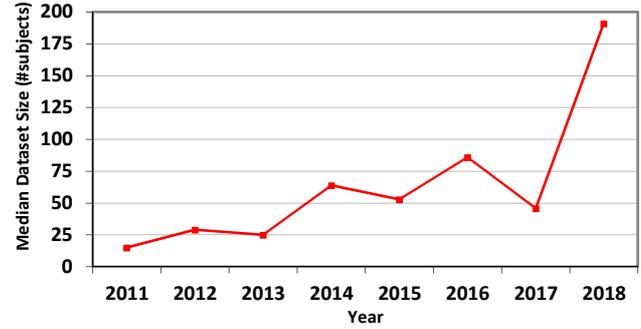

**FIGURE 3.** Median size of datasets used in MICCAI articles related to fMRI in each of the years 2011-2018.

The measures for CT are presented in Table 4, showing marked overall dataset growth over the study period. For CT, the graph of median dataset size per year is shown in Fig. 2.

The corresponding data for fMRI is presented in Table 5 and in Figure 3. The vertical scale in Figure 3 is different than in Figures 1 and 2, reflecting the larger growth rate.

## IV. STATISTICAL ANALYSIS AND PREDICTION

As a preliminary test of the dataset growth hypothesis, we first divided the range of years into two categories: 2011-2014 and 2015-2018. The distributions of the dataset sizes in the two categories were similar and non-normal. A Mann-Whitney U test was carried out to determine if there were differences in dataset sizes between the two time categories.

- For MRI, the median numbers of subjects for 2011-2014 (26) and 2015-2018 (55.5) were statistically significantly different, $U=19.115$, $p<0.001$.
- For CT, the median numbers of subjects for 2011-2014 (18.5) and 2015-2018 (36) were statistically significantly different, $U=8.311$, $p=0.006$.
- For fMRI, the median numbers of subjects for 2011-2014 (37) and 2015-2018 (77) were statistically significantly different, $U=10.493$, $p=0.003$.

These results corroborate the dataset-growth hypothesis.

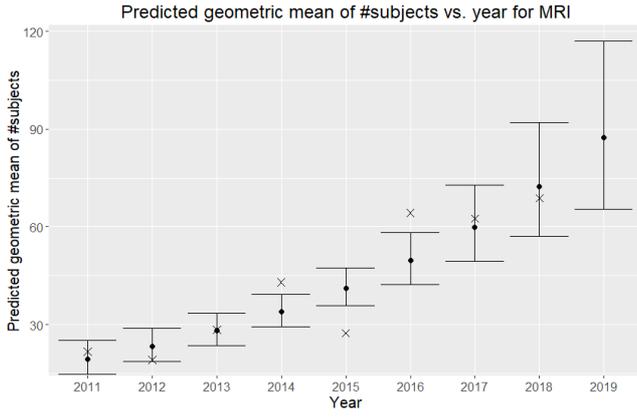

**FIGURE 4.** Predicted geometric mean (black dots) and confidence intervals of dataset sizes in MICCAI articles involving MRI for the years 2011-2019. The empirical geometric means (2011-2018) are shown for comparison (x marks), but note that the regression was based on the whole ensemble of MRI dataset sizes, not on the empirical geometric means.

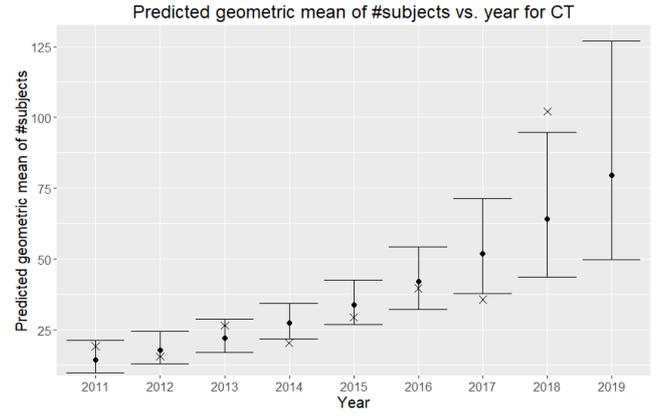

**FIGURE 5.** Predicted geometric mean (black dots) and confidence intervals of dataset sizes in MICCAI articles involving CT for the years 2011-2019. The empirical geometric means (2011-2018) are shown for comparison (x marks), but note that the regression was based on the whole ensemble of CT dataset sizes, not on the empirical geometric means.

Returning to the full year range 2011-2018, we used the natural logarithm ($ln$) transformation to normalize dataset sizes. Linear regression established that the year (after 2010) can statistically significantly predict the natural logarithm of dataset size.

For MRI, the model was statistically significant, $F(1,566)=38.720$, $p<0.001$. The model explained 6.2% (adjusted $R^2$) of the variance in the natural logarithm of dataset sizes. The year was statistically significant ($B=0.189$, CI=[0.129,0.249], $p<0.001$), where $B$ denotes slope and CI is its confidence interval. The regression equation is

$$\hat{E}(\ln N) = 2.771 + 0.189(y - 2010) \quad (1)$$

where $\hat{E}(\ln N)$ is the predicted mean of the natural logarithm of dataset sizes and $y$ is the year. Returning to the original scale of MRI dataset sizes, we obtain

$$\hat{G}(N) = e^{2.711+0.189(y-2010)} \quad (2)$$

Here, $\hat{G}(N)$ is the predicted *geometric mean* of MRI dataset sizes. The annual growth rate of the predicted geometric mean is about 21%, corresponding to ($e^{0.189} - 1$). Figure 4 shows $\hat{G}(N)$ with its confidence interval for each of the years 2011-2019. The empirical geometric means, taken from Table 3, are shown for comparison, but note that the regression is based on the whole ensemble of MRI dataset sizes, not on their empirical geometric means. The prediction for 2019 is of special interest, since MICCAI 2019 has not yet taken place at the time of writing. Specifically, we predict the geometric mean of MRI dataset sizes in MICCAI 2019 to be 87.5, with confidence interval [65.5,116.9].

The regression model was statistically significant for CT as well, $F(1,221)=21.273$, $p<0.001$. The model explained 8.4% (adjusted $R^2$) of the variance in the natural logarithm of dataset sizes. The year was statistically significant ($B=0.213$, [CI=0.122,0.305], $p<0.001$). The regression equation is

$$\hat{E}(\ln N) = 2.456 + 0.213(y - 2010) \quad (3)$$

Returning to the original scale of CT dataset sizes, we obtain

$$\hat{G}(N) = e^{2.456+0.213(y-2010)} \quad (4)$$

In Eq. (4), $\hat{G}(N)$ is the predicted geometric mean of CT dataset sizes. Here, the annual growth rate of the predicted geometric mean is about 24%. Figure 5 shows $\hat{G}(N)$ with its confidence interval for each of the years 2011-2019. The empirical geometric means, taken from Table 4, are shown for comparison, but note that the regression is based on the whole ensemble of CT dataset sizes, not on their empirical geometric means. We predict the geometric mean of CT dataset sizes in the upcoming MICCAI 2019 conference to be 79.6, with confidence interval [49.9,126.9].

Finally, for fMRI, the model was yet again statistically significant, $F(1,114)=27.130$, $p<0.001$. The model explained 18.5% (adjusted $R^2$) of the variance in the natural logarithm of dataset sizes. The year was statistically significant ($B=0.271$, CI=[0.168,0.374], $p<0.001$). The regression equation is

$$\hat{E}(\ln N) = 2.683 + 0.271(y - 2010) \quad (5)$$

Returning to the original scale of fMRI dataset sizes, we obtain

$$\hat{G}(N) = e^{2.683+0.271(y-2010)} \quad (6)$$

In Eq. (6), $\hat{G}(N)$ is the predicted geometric mean of fMRI dataset sizes. For fMRI, the annual growth rate of the predicted geometric mean is about 31%. Figure 6 shows $\hat{G}(N)$ with its confidence interval for each of the years 2011-2019. The empirical geometric means, taken from Table 5, are shown for comparison, but note that the regression is based on the whole ensemble of fMRI dataset sizes, not on their empirical geometric means. We predict the geometric mean of fMRI dataset sizes in the MICCAI 2019 conference to be 167.7, with confidence interval [104.9,268.0].

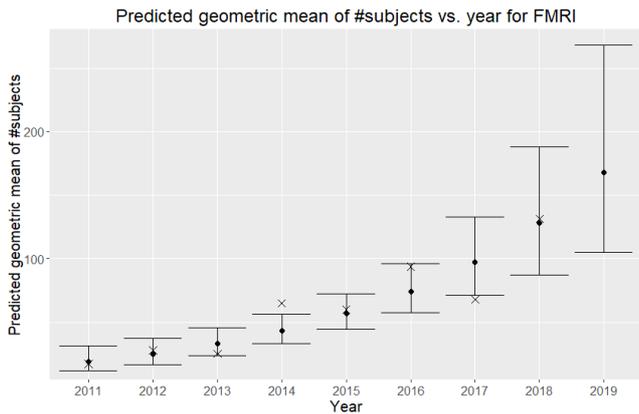

**FIGURE 6.** Predicted geometric mean (black dots) and confidence intervals of dataset sizes in MICCAI articles involving fMRI for the years 2011-2019. The empirical geometric means (2011-2018) are shown for comparison (x marks), but note that the regression was based on the whole ensemble of fMRI dataset sizes, not on the empirical geometric means.

## V. CONCLUSIONS

To our knowledge, this research is the first attempt to quantify the expectations of the research community regarding dataset size in medical image analysis research.

We scanned all the 2136 articles published in the MICCAI proceedings from 2011 to 2018. We extracted the dataset sizes from 907 papers relying on human data from three prevalent imaging modalities: MRI, CT and fMRI. Tables 3-5 and Figs. 1-3 describe the data. Human dataset size in MRI-related research nearly trebled from 2011 to 2018, while the CT and fMRI datasets grew at even faster rates. Similar trends are observed in the median, geometric mean and average values, though the average numbers are substantially higher than the median and geometric mean measures due to the occasional use of very large open datasets.

Statistical analysis using the Mann-Whitney U test corroborates the dataset growth hypothesis for all three modalities. Furthermore, regression analysis reveals statistically significant exponential growth in the geometric mean of the dataset sizes, albeit with large variability. The predicted annual growth rates in the geometric mean of the number of subjects in the datasets are about 21% for MRI, 24% for CT and 31% for fMRI. In slight analogy to Moore's law, these estimated growth rates can provide researchers, review boards and funding agencies with a tentative roadmap regarding dataset sizes in future medical image analysis studies.

In the upcoming MICCAI 2019 conference, we predict the geometric mean of the number of subjects in the datasets to be 87.5 with confidence interval [65.5,116.9] for MRI related articles, 79.6, with confidence interval [49.9,126.9] for CT and 167.7 with confidence interval [104.9,268.0] for fMRI.

The perception that "everyone participating in medical image evaluation with machine learning is data starved" [3] is not surprising given the exponentially increasing expectations regarding dataset size. Transfer learning [22] and data augmentation [23] are two popular and often successful strategies to alleviate the shortage of data, see [24]. Recently, Generative Adversarial Networks (GAN) have been used to create large sets of "fake" but credible new medical images based on a limited collection of genuine images, see e.g. [25]. If accepted by the research community, this strategy may bridge the gap between demand and supply of medical images for use in research and development.